\documentclass[11pt]{article}
\usepackage{pst-node,pst-tree,pst-char}
\def\itp#1{(\textit{#1}\/)}
\newtheorem{lemma}{Lemma}
\newtheorem{theorem}{Theorem}

\def\sqr{{\vcenter{\vbox{\hrule height.4pt%
	\hbox{\vrule width.4pt height5pt \kern5pt%
	\vrule width.4pt} \hrule height.4pt}}}}
\newcommand{\qed}{\hfill$\sqr$}
\newenvironment{proof}{\begin{trivlist}
\item[\hspace{\labelsep}{\bf\noindent Proof: }]
}{\hfill\qed\end{trivlist}}
\newsavebox{\boxtabbing}
{\end{tabbing}\end{minipage}\end{lrbox}%
\framebox[\columnwidth][l]{\usebox{\boxtabbing}}}
\newcommand{\doublespace}{\addtolength{\baselineskip}{.25\baselineskip}}
\newcommand{\remove}[1]{}
\def\insertop{\texttt{insert}}
\def\deleteop{\texttt{delete}}
\def\findop{\texttt{find}}
\def\locate{\texttt{locate}}
\def\intree{\texttt{intree}}
\def\root{\texttt{root}}
\def\free{\texttt{free}}
\def\curkey{\texttt{curkey}}
\def\curnode{\texttt{curnode}}
\def\scount{\texttt{count}}
\def\segment{\texttt{segment}}
\def\parent{\texttt{parent}}
\def\comma{\texttt{,}\;\;}
\def\high{\texttt{high}}
\def\low{\texttt{low}}
\def\key{\texttt{key}}
\begin{document}
\title{Available and Stabilizing 2-3 Trees}
\author{Ted Herman\thanks{This work is supported
by NSF CAREER award CCR-9733541.}
\\ University of Iowa \\ {\tt herman@cs.uiowa.edu}
\and
Toshimitsu Masuzawa\thanks{Supported in part by Japan Society
for the Promotion of Science (JSPS),
Grants-in-Aid for Scientific Research.
} \\
Graduate School of Engineering Science \\
Osaka University \\
1-3 Machikaneyama, Toyonaka 560-8531, Japan \\
{\tt masuzawa@ics.es.osaka-u.ac.jp}}
\date{\texttt{1 December 2000}}
\maketitle
\begin{abstract} \noindent
Transient faults corrupt the content and organization of data structures. 
A recovery technique dealing with such faults is stabilization,
which guarantees, following some number of operations on 
the data structure, that content of the data structure is legitimate.  
Another notion of fault tolerance is availability, 
which is the property that operations continue to be applied during the 
period of recovery after a fault, and successful updates are not
lost while the data structure stabilizes to a legitimate state. 
The available, stabilizing 2-3 tree supports 
\findop, \insertop, and \deleteop\
operations, each with $O(\lg n)$ complexity when the tree's state
is legitimate and contains $n$ items.  
For an illegitimate state, these operations have 
$O(\lg K)$ complexity where $K$ is the maximum capacity of the tree.
Within $O(t)$ operations, the state of the tree is guaranteed to
be legitimate, where $t$ is the number of nodes accessible via
some path from the tree's root at the initial state.
This paper resolves, for the first time, issues of dynamic 
allocation and pointer organization in a stabilizing data structure.
\end{abstract}
\doublespace
\section{Introduction}
Two important themes in the literature of fault tolerant design
are availability and self-repair.  A ``highly available'' system
continues to provide service (perhaps at degraded level) in spite
of failures of its components.  If component failures are transient,
then the system can repair the states of damaged components.  These
themes also apply to abstract data structures, which is useful for
object-oriented system design.  We call a data structure available
if each operation invocation returns a response consistent with
its effect on the data structure, in spite of arbitrary values in
all data structure fields (including pointers, keys, counters, and so on)
prior to the operation.  
We call a data structure stabilizing if, for any initial
state of the data structure, any sequence of operations applied
to the data structure brings it to a legitimate state. 
Separately, availability and stabilization have drawbacks:  
availability does not guarantee repair to a 
damaged structure and performance of 
operations can remain permanently degraded;  stabilization does not
make guarantees about the behavior of operations in the period
before repair has completed.  We therefore seek data 
structures that are available and stabilizing.  
\par
(Self-) stabilization is the topic of numerous investigations
in the field of distributed computing \cite{S93c}, but very few papers 
consider the question of stabilizing data structures.  The 
usual model for stabilizing algorithms is process-oriented,
meaning that variables subject to transient faults have 
dedicated processes that continually check and correct
faulty data.  We study passive data structures, for which the only
checking and correction occurs within the normal application of operations
(\findop, \insertop, \deleteop);  faults will not be corrected unless
operations are applied.  
\par
Related to this work are papers such as \cite{FMRT96} which 
consider transient corruption of one portion of data, but rely on 
control variables that initiate computation.  Our assumption is that 
the data structure may be damaged, but each operation starts 
cleanly with its internal control variables uncorrupted.  
We wish to constrain the behavior of operations in cases where data 
is faulty by an availability guarantee, which resembles previous
work on graceful degradation \cite{HW94}.  With the exception of
a few recent papers \cite{UKMF97,DH97,H99}, most stabilizing algorithms
do not constrain behavior during periods while data is corrupt.  
Moreover, the stabilization time of our construction is adaptive:
the stabilization time depends on the size of the initial (possibly
damaged) tree structure, and in this respect our research follows 
a recent trend of adaptive stabilization times \cite{AD97a,KS97}.  
\par
Our contribution is a new form of stabilizing data structure,
which constrains behavior of every operation, brings the 
the data structure to a legitimate state over a sequence of operations, 
and does so with an adaptive stabilization time. 
This paper goes beyond our previous investigation of heaps \cite{HM00}
by showing how availability and stabilization are possible for 
a dynamic data structure using pointers.
\section{Stabilizing Search Tree Specification}

The construction presented in this paper is one type of 
data structure supporting \findop, \insertop, and \deleteop\ operations
with logarithmic running times.  The behavior of operations
and the specification of stabilization properties are 
general, and we state them here in terms of a generic 
\emph{search tree}.  In this section, the behavior of a search
tree is initially specified without considering stabilization 
properties.  Subsequently this specification is revised to 
include stabilization criteria.  

\subsection{Search Tree Operations} \label{searchops}        

A search tree is an associative memory containing \emph{items} of 
the form $\langle\textit{key},\textit{datum}\rangle$. 
The \emph{capacity} $K$ of the search tree is an upper bound on the 
number of items that the search tree may store. 
Let $\cal H$ be an infinite sequential 
history of operations on a search tree. 
Each operation consists of a pair $\langle\textit{inv},\textit{resp}\rangle$
where the invocation \textit{inv} is one of \{\findop,\insertop,\deleteop\} 
accompanied by calling parameters, 
and the response \textit{resp} is as follows:
for a \findop\ or \deleteop\ invocation, the response is either 
``missing'' or an item;  for an \insertop\ invocation,
the response is either ``ack'' or ``full''.                
The complete signature for an \insertop\ invocation is 
$\insertop(\textit{key},\textit{datum})$, however to streamline
the presentation we ignore the \textit{datum} component and 
write $\insertop(\textit{key})$ in subsequent sections.
The signatures for the other invocations are 
$\findop(\textit{key})$ and $\deleteop(\textit{key})$.
We say an \insertop\ invocation \emph{succeeds}
if its response is ``ack'' and \emph{fails} if its response is 
``full''; similarly, a \findop\ or \deleteop\ invocation is 
said to fail if its result is ``missing'' and otherwise is successful.

Semantics of operations are given in terms of the content of the
search tree, which we describe using the operation history.
The search tree \emph{content} is defined for any point between
operations in a history $\cal H$.  For completeness, we define the
content before any operation in $\cal H$ to be the empty set
(the search tree initially contains no items).  Let $t$ be 
a point in $\cal H$ between operations;  the content of the
search tree at point $t$, denoted by $C_t$, is the bag
of items $C_t = I_t \setminus D_t$, where $I_t$ is the bag of
items successfully inserted prior to point $t$, and $D_t$ is
the bag of items successfully deleted prior to point $t$.

Search tree operations satisfy the following constraints:
\itp{1} a $\deleteop(k)$ ($\findop(k)$) 
invocation immediately following any point
$t$ in any history returns ``missing'' iff there exists no $d$ such that 
$\langle k,d\rangle\in C_t$, and otherwise returns some item
$\langle k,d\rangle\in C_t$; and \itp{2} an $\insertop(k,d)$ operation   
immediately following any point $t$ fails iff $|C_t|\geq K$, 
and otherwise returns ``ack''.  
From \itp{1} and \itp{2} one can show intuitive 
search tree properties, for instance, 
no \findop\ returns an item not previously inserted. 
A \emph{balanced} search tree satisfies additional constraint, 
also specified with respect to any point $t$ in a history: \itp{3}
the running time of any operation immediately following $t$ 
is $O(\lg |C_t|)$.  

\subsection{Available and Stabilizing Operations} \label{avstable}

Transient faults inject arbitrary data
into data structures, which is modeled in the 
literature of stabilizing algorithms by considering 
arbitrary initial states --- the state following a 
transient fault is the ``initial'' state for subsequent
computation.  \S\ref{searchops}'s characterization of 
search tree behavior depends on $C_z$ being empty at 
the initial point $z$ in any history, so we consider 
next a characterization admitting arbitrary initial
content in a search tree. 

Let $\cal P$ denote a history fragment, starting
from an initially empty search tree, that consists entirely of successful
\insertop\ operations.  To specify behavior of $\cal H$ 
for an arbitrary initial
search tree, let ${\cal H}' = {\cal P}\circ{\cal H}$.   
A search tree implementation is \emph{available} if for 
every history $\cal H$ of operations 
there exists ${\cal P}$ such that 
${\cal H}'$ satisfies \itp{1} for all operations;
and \itp{2'} no \insertop\ operation at any point $t$ succeeds
if $|C_t|\geq K$.  A balanced search tree implementation is available
if it is an available search tree and the running
time of every operation is $O(\lg K)$. 

Note that \itp{2} is not required for availability:
an \insertop\ operation at a point $t$ is allowed to fail 
when $C_t<K$.  A trivial implementation of an available 
search tree would be one that fails all \insertop\ operations.
Although this definition of availability weakens the specification,
it does provide safety guarantees for the search tree content.
For instance, if $\insertop(k,d)$ does succeed, any subsequent
$\findop(k)$ will succeed at least until a $\deleteop(k)$ operation
is applied.  

Let ${\cal H}_v$ denote the suffix of a history $\cal H$
following a point $v$ in $\cal H$.  
A search tree implementation is \emph{stabilizing} if for 
every history $\cal H$ of operations there exists a point $v$ and
a history fragment $P$ such that all operations in 
${\cal P}\circ{\cal H}_v$ satisfy \itp{1} and \itp{2}.   
A balanced search tree implementation is stabilizing if
it is a stabilizing search tree implementation 
and every operation in ${\cal H}_v$ satisfies \itp{3}.

The point $v$ in the definition of 
stabilization divides the history $\cal H$
into illegitimate and legitimate parts.  Prior to $v$, the 
content of the search tree has no relation to the responses
of invocations;  the behavior could be chaotic in this portion
of the history.  Following $v$, all operations behave normally
with respect to some ``initializing'' history $\cal P$.  
A possible implementation of a stabilizing search tree would
be one that, after some number of operations, 
resets the content of the search
tree to the empty set ($\cal P$ would be empty in this case). 
The portion of history $\cal H$ prior to point $v$ is called
the \emph{convergence period}, and the worst case number of
operations in the convergence period, 
taken over all possible histories for a 
stabilizing search tree implementation, is
the \emph{stabilization time} of the implementation. 

Availability alone does not guarantee progress, since 
\insertop\ operations can continue to fail in a history
although the search tree is not full.
Stabilization does guarantee progress eventually, 
but items that have been inserted successfully during
the convergence period could be lost before the search
tree stabilizes.  We therefore aim for a search tree
that is both available and stabilizing.  For example, 
a balanced, available, stabilizing, search tree enjoys safety
and timing guarantees throughout any history 
(each operation has $O(\lg K)$
running time and the semantics of invocation responses
are well-defined) and after convergence, behavior
is what one expects of a balanced search tree. 
\section{Construction for Stabilization} \label{overview}
Our stabilizing search tree is a modification of 
a conventional 2-3 tree implementation.  After
briefly reviewing this basic 2-3 tree, this section 
surveys the technical enhancements introduced for stabilization
and availability.   

\subsection{2-3 Tree Review}

\begin{figure}[ht]
\begin{minipage}{\columnwidth}
\vspace{3ex}\par
\begin{pspicture}(-3,-4.5)
\psset{levelsep=50pt,treesep=0.2cm}
\pstree{\Tr{\psframebox{$\begin{array}{c}157\\230\end{array}$}}}{
  \pstree{\Tr{\psframebox{$\begin{array}{c}120\\141\end{array}$}}}{
     \Tcircle{120}\Tcircle{141}\Tcircle{157}
     }
  \pstree{\Tr{\psframebox{$\begin{array}{c}188\\205\end{array}$}}}{
     \Tcircle{188}\Tcircle{205}\Tcircle{230}
     }
  \pstree{\Tr{\psframebox{$\begin{array}{c}240\\252\end{array}$}}}{
     \Tcircle{240}\Tcircle{252}
     }
  }
\end{pspicture}
\end{minipage}
\caption{typical 2-3 tree}
\label{tree-one}
\end{figure}
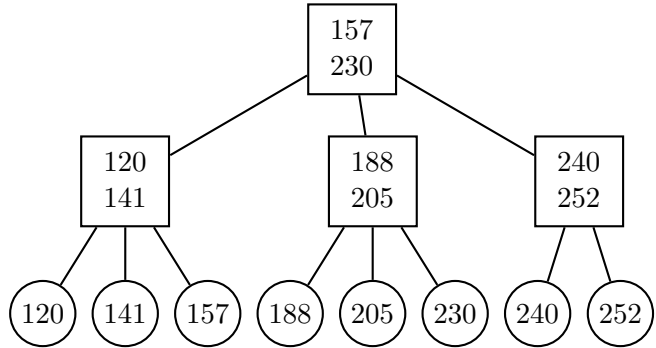

A 2-3 tree is a balanced search tree with the following 
structure \cite{AHU74}:  each non-leaf node has either two or three
children and the path length from root to leaf is the same
for every leaf.  Each leaf contains one item, and non-leaf
nodes contain one or two keys of items in their subtrees.
Figure \ref{tree-one} presents an example 2-3 tree, showing
how interior nodes have the maximum keys of their two left-most
subtrees.  It follows from this definition that a 2-3 tree of
height $h$ contains between $2^h$ and $3^h$ items. 

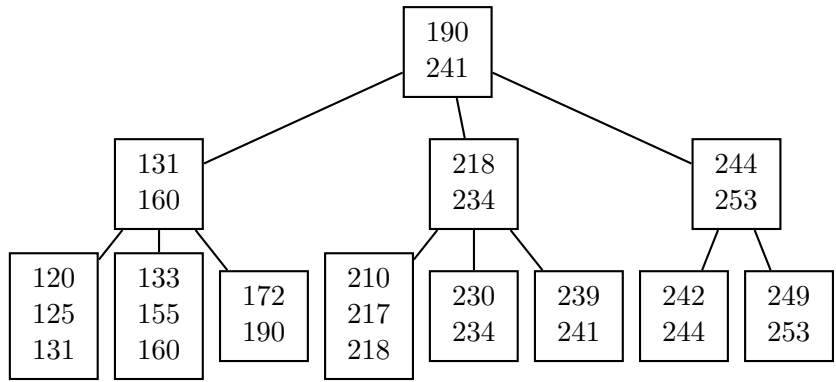
\begin{figure}[ht]
\begin{minipage}{\columnwidth}
\vspace{4ex}\par
\begin{pspicture}(-3,-4.5)
\psset{levelsep=50pt,treesep=0.2cm}
\pstree{\Tr{\psframebox{$\begin{array}{c}190\\241\end{array}$}}}{
  \pstree{\Tr{\psframebox{$\begin{array}{c}131\\160\end{array}$}}}{
     \Tr{\psframebox{$\begin{array}{c}120\\125\\131\end{array}$}}
     \Tr{\psframebox{$\begin{array}{c}133\\155\\160\end{array}$}}
     \Tr{\psframebox{$\begin{array}{c}172\\190\end{array}$}}
     }
  \pstree{\Tr{\psframebox{$\begin{array}{c}218\\234\end{array}$}}}{
     \Tr{\psframebox{$\begin{array}{c}210\\217\\218\end{array}$}}
     \Tr{\psframebox{$\begin{array}{c}230\\234\end{array}$}}
     \Tr{\psframebox{$\begin{array}{c}239\\241\end{array}$}}
     }
  \pstree{\Tr{\psframebox{$\begin{array}{c}244\\253\end{array}$}}}{
     \Tr{\psframebox{$\begin{array}{c}242\\244\end{array}$}}
     \Tr{\psframebox{$\begin{array}{c}249\\253\end{array}$}}
     }
  }
\end{pspicture}
\end{minipage}
\caption{2-3 tree with implicit leaves}
\label{tree-two}
\end{figure}

This definition of a 2-3 tree differentiates between
item nodes (leaves) and interior nodes, which is a detail
not important for our presentation.  In subsequent discussion
the leaves of 2-3 trees are omitted; to show the keys of 
all items, parents of the omitted leaves list the key values
of all their children.  Figure \ref{tree-two} is an example
with leaves omitted while all item keys are shown.  Another
interpretation of this representation is that two or three
items are contained in each leaf of a tree of $n>1$ items.

Operations on a 2-3 tree of $n$ items (\findop, \deleteop, \insertop) 
have $O(\lg n)$ running time because tree height is logarithmic.
The \findop\ operation has a straightforward implementation;  
the \insertop\ and \deleteop\ operations rearrange keys within
interior nodes, possibly splitting nodes, merging nodes, or
reassigning the root node location and adjusting tree height
to maintain the invariant that each interior node have two or three 
children.  

Our definition of availability depends on a maximum capacity
$K$ for a balanced search tree.  Many texts describe tree 
operations in detail, but few consider the case of a 2-3 
tree with fixed capacity.  For simplicity of presentation, 
we suppose that $K$ is a power of three, and fix the maximum
path length from root to leaf at $\textit{pmax}=\log_3 K$.
A practical difficulty in defining capacity as a fixed 
threshold $K$ for deciding success or failure of an 
\insertop\ operation is that a set of items can have 
numerous representations as a 2-3 tree, some of which 
make insertion harder than others. 

For instance, no \insertop\ operation 
can succeed if it would increase the
tree height beyond \textit{pmax}, 
and requirement \itp{3} for logarithmic running
time precludes extensive key redistribution by any single
\insertop\ operation.  Consider the case of $\textit{pmax}=3$
(so $K=27$) and the tree of Figure \ref{tree-two}.  
A conventional implementation of $\insertop(123)$ in this
case would result in a recursive node split and increase in tree height;
we make the design decision that $\insertop(123)$ should fail 
here, although the tree has only 19 items.  Formally, the 
consequence of this design decision is that definitions of 
2-3 tree capacity and condition \itp{2} for the success
of an \insertop\ operation should be revised;  to 
simplify the presentation we omit this level of detail.

The operation definitions in \S\ref{searchops} 
refer to the content of a search tree as a bag of items,
which allows for duplicate items in a search tree.
Duplicate items in a 2-3 tree are usually differentiated
by extending the key with a sequence number or a node  
address.  We omit presenting details of this standard technique.

\subsection{Structural Modifications for Stabilization} \label{structmod}

The literature of stabilizing algorithms is primarily oriented
to distributed computing, in which a recurrent theme is establishing 
some global system property using processes with only local resources
and limited communication facilities.  Not surprisingly, the 
technique of many stabilizing algorithms is:  processes detect
illegitimate global states by local checking, and thereafter 
effect state correction either locally or by initiating a system 
reset.  The problem of stabilizing 2-3 trees is not distributed,
but shares some characteristics with distributed systems:  the 
legitimacy of the data structure is a global property, but 
at most $O(\lg K)$ nodes can be visited by any single tree 
operation.  We therefore follow some traditional stabilization
techniques, starting with local checking to determine legitimacy
of data.  After describing some of the challenges posed by 
illegitimate data, we describe modifications to conventional
2-3 tree representations so that local checking is enabled.  

We begin by considering how a transient fault can 
disrupt the content and organization of 2-3 tree.  
Such disruption has an impact on three entities, 
keys, pointers, and auxiliary variables.  Key corruption
violates the condition that each internal node key has 
the maximum item key value of the corresponding subtree.
After a fault, key values can be duplicated and out of order.
Pointer corruption damages the tree structure, possibly producing
orphan nodes, ancestry cycles, and references to arbitrary 
locations in memory.  Auxiliary variable corruption can 
cause the location of the root node to be lost, can invalidate
counters, and damage the mechanism of node allocation and 
deallocation.  

The standard implementation of a 2-3 tree uses $(k-1)$ keys 
in a node with $k$ children.  Our first modification is to give
each node the same number of keys as it has children, and to 
strengthen navigation by using a pair of keys $(\low_p,\high_p)$ 
for each child $p$.  The value $\low_p$ is a lower bound on 
the minimum item value of the subtree rooted at $p$, 
and the value $\high_p$ is an upper bound on the 
maximum key value of the subtree rooted at $p$.  Thus each
node has minimum and maximum bounds for all its children and their
subtrees.  This modification is a first step to local checking:
it is now possible to verify key values by comparison between
parent and child.  The only keys remaining unverifiable are
those of the items, which have no children.  With this modification,
each key pair within a node has a associated child pointer.  If the
pointer associated with a key is null, then that key is called
\emph{irrelevant}.  Only relevant keys and their corresponding
pointers are verified by local checking.     
We use the notation $\key_p$ as shorthand for $(\low_p,\high_p)$.  
  
Two modifications enable local detection of pointer corruption.
The first is to support each tree link with a double pointer.
Each node except the root now has a pointer to its parent.  
This provides a ``sanity'' check for pointers, so that a 
child pointer can be checked by comparing fields in two nodes: 
the pointer check, for relevant $\key_q$ in node $p$, consists of 
testing the equality $\parent(q)=p$. 
The situation of multiple parents of one node is now easily
identified.  However some global properties are not verified 
by this modification, including the property that every path
from root to leaf should have the same length.  The second 
modification addressing pointer corruption is to use a static
allocation scheme for the placement of nodes in memory. 

The storage used for allocation of tree nodes is partitioned into
\textit{pmax} segments labeled $S_i$, $0<i\leq pmax$.   
Node placement in these segments invariantly satisfies:
a node $p$ at height $i$ in the 2-3 tree resides in 
segment $S_i$ (for completeness, we can let $S_0$ be the segment
containing the data items in the tree). 
This constraint enables simple and local ``type checking''
of child and parent pointers:  each child (parent) pointer of
a node in segment $S_i$ should refer to a node within segment
$S_{(i-1)}$ ($S_{(i+1)}$).  For a node $p$, let 
$\segment(p)$ denote the segment containing $p$.    

The partition of storage into segments $\{\,S_i\;|\; 0<i\leq pmax\}$
dictates that allocation and deallocation of nodes occur
within each segment.  Each segment therefore has a free list 
of unallocated nodes, which is managed as a stack:  a newly 
deallocated node is added to the front of the free list, 
and allocation consists of removing the first node in the free 
list.  Invariantly, every ``next'' pointer of a node in
free list of $S_i$ should either be null or refer 
to some node within $S_i$.  Each node in $S_i$ should either
be a tree node or an element of the free list.  The total
number of nodes in $S_i$ is denoted $|S_i|$.  For $i>1$, 
$|S_i|=1+\lceil S_{i-1}/2\rceil$, and $|S_1|=\lceil K/2\rceil$. 

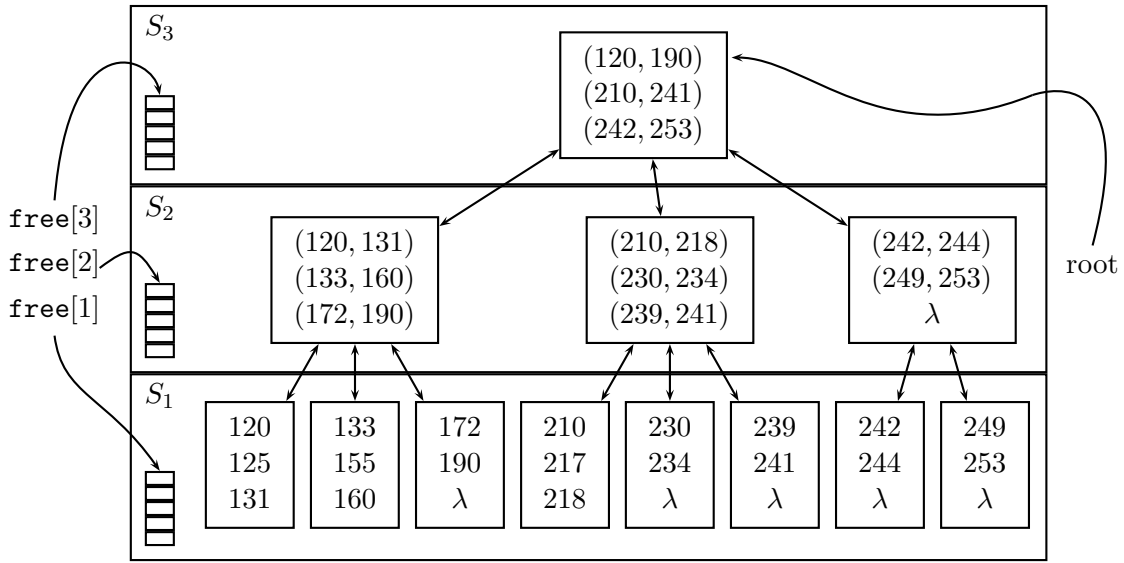
\begin{figure}[ht]
\begin{minipage}{\columnwidth}
\vspace*{7ex}\par
\begin{pspicture}(-3,-5.5)
\psframe(-1,-1.2)(11.2,1.2)
\psframe(-1,-3.7)(11.2,-1.2)
\psframe(-1,-6.2)(11.2,-3.7)
\multips(-0.8,-1)(0,0.2){5}{\psframe(0,0)(0.4,0.2)}
\multips(-0.8,-3.5)(0,0.2){5}{\psframe(0,0)(0.4,0.2)}
\multips(-0.8,-6)(0,0.2){5}{\psframe(0,0)(0.4,0.2)}
\rput(-2,-1.65){\free[3]}
\rput(-2,-2.25){\free[2]}
\rput(-2,-2.85){\free[1]}
\psbezier{->}(-2,-1.4)(-1.8,1)(-0.7,0.5)(-0.6,0)
\psbezier{->}(-1.4,-2.3)(-1.0,-2.0)(-0.9,-2.0)(-0.6,-2.5)
\psbezier{->}(-2,-3.2)(-1.9,-4)(-1.3,-4)(-0.6,-5)
\rput(-0.6,0.9){$S_3$}
\rput(-0.6,-1.5){$S_2$}
\rput(-0.6,-4){$S_1$}
\rput(11.8,-2.25){root}
\psbezier(11.8,-2)(12.4,0.3)(11.3,0.1)(11,0)
\psbezier{->}(11,0)(9,-0.8)(8,0.5)(7,0.5)
\psset{levelsep=70pt,treesep=0.2cm,arrows=<->}
\pstree{\Tr{\psframebox{$\begin{array}{c}
		(120,190)\\(210,241)\\(242,253)\end{array}$}}}{
  \pstree{\Tr{\psframebox{$\begin{array}{c}
		(120,131)\\(133,160)\\(172,190)\end{array}$}}}{
     \Tr{\psframebox{$\begin{array}{c}120\\125\\131\end{array}$}}
     \Tr{\psframebox{$\begin{array}{c}133\\155\\160\end{array}$}}
     \Tr{\psframebox{$\begin{array}{c}172\\190\\\lambda\end{array}$}}
     }
  \pstree{\Tr{\psframebox{$\begin{array}{c}
		(210,218)\\(230,234)\\(239,241)\end{array}$}}}{
     \Tr{\psframebox{$\begin{array}{c}210\\217\\218\end{array}$}}
     \Tr{\psframebox{$\begin{array}{c}230\\234\\\lambda\end{array}$}}
     \Tr{\psframebox{$\begin{array}{c}239\\241\\\lambda\end{array}$}}
     }
  \pstree{\Tr{\psframebox{$\begin{array}{c}
		(242,244)\\(249,253)\\\lambda\end{array}$}}}{
     \Tr{\psframebox{$\begin{array}{c}242\\244\\\lambda\end{array}$}}
     \Tr{\psframebox{$\begin{array}{c}249\\253\\\lambda\end{array}$}}
     }
  }
\end{pspicture}
\vspace*{3ex}\par ~
\end{minipage}
\caption{2-3 tree within segments}
\label{tree-three}
\end{figure}

Auxiliary to the segments, the following pointers are needed:
a pointer \root\ to the root node, and \textit{pmax} additional pointers,
which start the free lists of the segments.  Figure \ref{tree-three}
summarizes the structure of segments and auxiliary pointers
as applied to the previous example of Figure \ref{tree-two}.  
The symbol $\lambda$ indicates a null pointer (which makes the
corresponding key value irrelevant).  

\section{Operation Modifications for Stabilization} \label{modops}

Operations on a 2-3 tree are explained in many texts, 
and we suppose the reader is familiar with some conventional 
implementation of these operations, including node splits,
merges, and recursive cases for these events.  To simplify
the presentation, we omit a full description of the operations
and concentrate on the modifications needed for stabilization. 
For stabilization, \findop, \deleteop, and \insertop\ operations use 
local checking to detect illegitimate conditions in the 
2-3 tree and make corrections bringing the data structure
to a legitimate state.

Two themes of the modifications to operations are
truncation of the 2-3 tree to a legitimate fragment
and background processes that reorganize the data structure.
Our basic design decision is to trust key and pointer data 
from the tree's root downward, which has the consequence that
an item in a 2-3 tree will be lost if a transient fault damages
the path leading to that item.  The rationale for this
decision is due to the definition in the next subsection.
We return in \S\ref{conclusion} to discuss further the 
issues of data loss due to transient faults.

\subsection{The Active Tree}

Given arbitrary initial (and possibly corrupt) values for 
variables and storage, many of the key and pointer properties
of the 2-3 tree described in \S\ref{structmod} could be falsified.
Nevertheless, it is possible in many situations to identify some
fragment, starting from the root, which enjoys properties of 
of a search tree.

A \emph{state} of the 2-3 tree is a specification of the values
for all variables and storage used by the data structure.  
The \emph{active tree} is defined with respect to a given state
$\sigma$.  The definition of the active tree is recursive, depending
on an intermediate tree called the \emph{base tree}, denoted by $T_\sigma$. 
The items in the active tree define the data structure content;
in proofs of availability and stabilization, the items of the initial
active tree define the initial sequence $\cal P$ of successful
\insertop\ operations.

Let $T_{\sigma}$ be, for a state $\sigma$, the
tree defined as follows.  If $\root=\lambda$ or 
$\segment(\root)\not\in[1,\textit{pmax}]$, then $T_{\sigma}$ is
empty;  otherwise \root\ specifies the root node of $T_{\sigma}$.
The remaining nodes of $T_{\sigma}$ are defined recursively:
if $p\in T_{\sigma}$, and $p$ has a relevant key $\key_q$
such that $\parent(q)=p$ and $\segment(q)=\segment(p)-1$,
then $q\in T_{\sigma}$.  

The active tree is obtained by 
applying the following rules, as many times as possible,
to the base tree (initially, let $T=T_{\sigma}$), giving
priority to rule application in higher level segments
over lower level segments, and giving priority to 
rules in the order listed where more than one rule is
applicable to a particular node.  
\begin{quote}
\begin{trivlist}
\item \itp{a}  if $p\in T$ has a key $(\low_q,\high_q)$ with
$q\in T$ such that $\high_q\leq\low_q$, then remove $q$ and 
its descendants from $T$.
\item \itp{b}  if $p\in T$ has a key $\key_q$ with 
either $\segment(q)=1$ and $q$ has no keys, or 
$\segment(q)>1$ and $q$ has no relevant keys, 
then remove $q$ and its descendants from $T$.
\item \itp{c}  if $p\in T$ has two relevant keys 
$(\low_q,\high_q)$ and $(\low_r,\high_r)$ that overlap ranges 
(such as $\high_q>low_r$), then one of these relevant keys is
made irrelevant by removing its associated child and all its
descendants from $T$.  The key to be made irrelevant
is some deterministic choice; furthermore, if 
more than one key pair overlap ranges, the choice
of which overlap to resolve is also deterministic
(such determinism is necessary for a unique definition of 
the active tree).
\item \itp{d}  if $p\in T$ has a key $\key_q$
for $q\not\in T$, then remove $q$ and all its descendants from $T$.
\item \itp{e}  if $p\in T$ has a relevant key 
$(\low_q,\high_q)$ and $q\in T$ has a relevant key 
$\key_r$ outside the range $(\low_q,\high_q)$, then  
remove $r$ and its descendants from $T$.
\end{trivlist}
\end{quote}
The intuition of \itp{a}--\itp{e} is that keys at greater
height in $T_\sigma$ are more trustworthy than lower ones.
So if a child has a key not reflected by its parent's
key range for that child, some (or all) of the child's
keys should be made irrelevant in the active tree. 

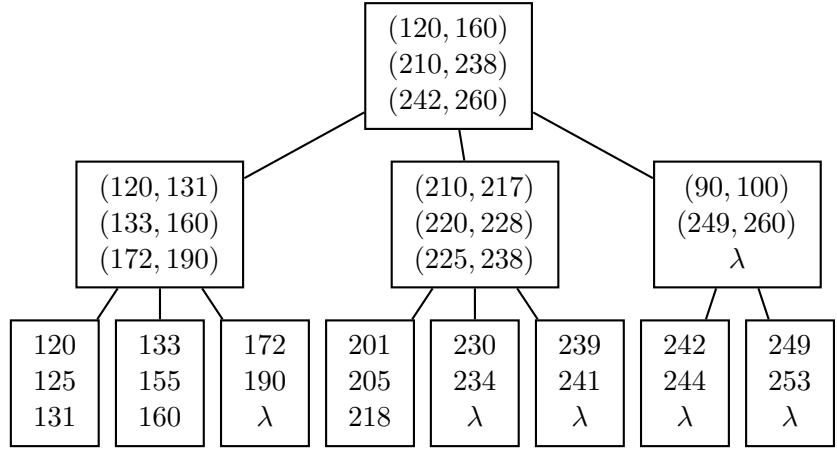
\begin{figure}[ht]
\begin{minipage}{\columnwidth}
\vspace{4ex}\par
\begin{pspicture}(-3,-4.5)
\psset{levelsep=60pt,treesep=0.2cm}
\pstree{\Tr{\psframebox{$\begin{array}{c}
	(120,160)\\(210,238)\\(242,260)\end{array}$}}}{
  \pstree{\Tr{\psframebox{$\begin{array}{c}
	(120,131)\\(133,160)\\(172,190)\end{array}$}}}{
     \Tr{\psframebox{$\begin{array}{c}120\\125\\131\end{array}$}}
     \Tr{\psframebox{$\begin{array}{c}133\\155\\160\end{array}$}}
     \Tr{\psframebox{$\begin{array}{c}172\\190\\\lambda\end{array}$}}
     }
  \pstree{\Tr{\psframebox{$\begin{array}{c}
	(210,217)\\(220,228)\\(225,238)\end{array}$}}}{
     \Tr{\psframebox{$\begin{array}{c}201\\205\\218\end{array}$}}
     \Tr{\psframebox{$\begin{array}{c}230\\234\\\lambda\end{array}$}}
     \Tr{\psframebox{$\begin{array}{c}239\\241\\\lambda\end{array}$}}
     }
  \pstree{\Tr{\psframebox{$\begin{array}{c}
	(90,100)\\(249,260)\\\lambda\end{array}$}}}{
     \Tr{\psframebox{$\begin{array}{c}242\\244\\\lambda\end{array}$}}
     \Tr{\psframebox{$\begin{array}{c}249\\253\\\lambda\end{array}$}}
     }
  }
\end{pspicture}
\vspace{3ex}\par
\end{minipage}
\caption{illegitimate keys in a 2-3 tree}
\label{tree-four}
\end{figure}

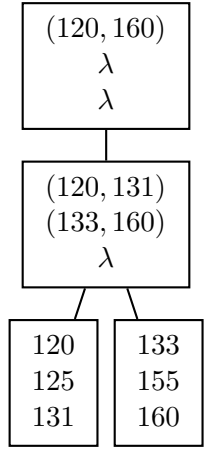
\begin{figure}[ht]
\begin{minipage}{\columnwidth}
\vspace{4ex}\par
\begin{pspicture}(-6,-4.5)
\psset{levelsep=60pt,treesep=0.2cm}
\pstree{\Tr{\psframebox{$\begin{array}{c}
	(120,160)\\\lambda\\\lambda\end{array}$}}}{
  \pstree{\Tr{\psframebox{$\begin{array}{c}
	(120,131)\\(133,160)\\\lambda\end{array}$}}}{
     \Tr{\psframebox{$\begin{array}{c}120\\125\\131\end{array}$}}
     \Tr{\psframebox{$\begin{array}{c}133\\155\\160\end{array}$}}
     }
  }
\end{pspicture}
\vspace{3ex}\par
\end{minipage}
\caption{active tree of the illegitimate 2-3 tree}
\label{tree-five}
\end{figure}

Figure \ref{tree-four} shows an example of a 2-3 tree with illegitimate
key values.  The example has several violations
of expected 2-3 tree properties:  the root does not contain the 
maximum key values of all its children;  the key value 228 is smaller
than any item in the corresponding subtree;  the key value 217 in
$S_2$ does not equal the maximum key value of the corresponding child. 
After applying rules \itp{a}--\itp{e}, the active tree
pictured in Figure \ref{tree-five} results (assuming the key
values at level $S_1$ of Figure \ref{tree-four} are legitimate).  

\subsection{Truncation} \label{truncate}

The structural modifications introduced in \S\ref{structmod} enable
each operation execution to perform extra measures of local checking
and correcting without increasing the operation's time complexity.
\emph{Truncation} is one step in local correction:  it assigns 
$\lambda$ to selected pointers, bringing the tree closer to the definition
of an active tree.  \emph{Node truncation}, 
for a node $p$, is the following procedure.  
\begin{quote}
If $\segment(p)=0$, there are no changes to $p$;  otherwise,
\begin{enumerate}
\item while a rule \itp{a}--\itp{d} is applicable to some relevant 
key $\key_q$ in $p$, make $\key_q$ irrelevant by assigning $\lambda$ to the 
associated pointer for each $q$ that is to be removed by
\itp{a}--\itp{d}; 
\item for each relevant key $\key_q$ in $p$, apply rule \itp{e}
as many times as needed to remove relevant keys from $q$;
if $q$ has no relevant keys as a result of this step, then
apply rule \itp{b} to make $\key_q$ irrelevant by assigning
$\lambda$ to its associated pointer.
\end{enumerate}
\end{quote}
The node truncation procedure is applied in both preorder and postorder
senses of visitation by each operation.  That is, \findop\ (\insertop, 
\deleteop) applies node truncation before examining or processing 
a node, in root to leaf order.  The preorder application of truncation
ensures that the operation does not follow paths outside the active tree.
Postorder application of node truncation speeds up stabilization.  Because
truncations at lower levels can occur after the preorder processing of
a given node, a postorder repetition of the node truncation procedure may
result in additional changes bringing the tree to its active form. 
Note that the time overhead of node truncation is $O(1)$,
so truncation does not increase the time complexity of operations.

\subsection{Merge and Collapse} \label{collapse}

Truncation brings a tree with illegitimate keys and pointers
to the form of an active tree.  However as Figure \ref{tree-five} shows,
active trees may not be 2-3 trees, because there may be single-child
nodes. Two procedures convert such an unbalanced tree to a 2-3 tree:
merging ``only child'' nodes with siblings and collapsing the root.
The second modification for local checking and correction is to 
have \findop, \insertop, and \deleteop\ apply a merge and collapse
procedure after truncation.

The \emph{merge and collapse} procedure deals with a node $p$ 
(after applying the truncation procedure) that has a single child.  
There are three cases for $p$:
\begin{enumerate}
\item if $p$ is the root with only child $q$, then assign 
$\root\leftarrow q$ (this is a \emph{collapse}); 
\item if $p$ is not the root, $p$ has parent $s$, 
and $p$ has a sibling $r$ such that
$r$ contains at most two relevant keys, then merge nodes $p$ and $r$ 
and adjust keys and pointers of $s$ appropriately;
\item if $p$ is not the root, $p$ has parent $s$, 
and $p$ has a sibling $r$ with three
relevant keys, then move one key $k$ (and its associated child) from $r$
to $p$ and adjust keys of $s$ appropriately (the choice of $k$ will be
the largest or smallest key of $r$, depending on the sibling relation
between $p$ and $r$). 
\end{enumerate}
Cases 2 and 3 are not exclusive; in a situation where both 
cases exist, some choice (deterministic or nondeterministic) is acceptable
to implement the procedure.    

Each operation on the data structure, after applying truncation,  
then applies the merge and collapse procedure.  Since the overhead
for merge is $O(1)$, the time complexity of an operation does not
increase due to the merge procedure. 

\subsection{Node Allocation}

A successful \insertop\ operation increases the number of nodes
in the tree and may increase the height as well.  While conventional
2-3 implementations allocate and deallocate nodes, the modifications
we introduce are to allocate on a segment basis and to locally check 
nodes on a free list to verify their availability. 

We describe the allocation scheme in terms of an array of 
node structures for each segment.  Auxiliary to each segment $S_i$,
let $\free[i]$ point to the head of the 
free list of unallocated nodes (as illustrated in Figure \ref{tree-three}). 
We make the following convention:
$p$ is \emph{detached} iff $p$ is in the free list, 
\root\ does not point to $p$, and $\parent(p)=\lambda$.    
Node allocation can occur at $S_i$ iff $\free[i]=p$ where 
$p$ is a detached node within segment $S_i$. 
Thus it is not sufficient for a node to appear on a free 
list for it to be detached --- the local check verifies
$\parent(p)=\lambda$.  The \emph{size} of a free
chain of $S_i$ is defined to be the number of detached nodes, counting
from $S_i$'s free chain pointer, until either a (next) pointer
leads outside of $S_i$ or leads to a node that is not detached.  
The size of $S_i$'s free chain is denoted by $f_i$.

A precondition for successful insertion is that sufficiently many 
nodes are detached, so an \insertop\ operation must test for detached 
nodes to determine whether the operation will succeed or fail.  
A simple implementation of such a test is to provide one extra 
node (beyond what is required for the capacity $K$) in each segment.
An \insertop\ then fails if, at any level from the root down to 
$S_1$, there are no detached nodes.  This test takes $O(\ell)$ time,
where $\ell$ is the height of the tree.   

Node deallocation occurs due to \deleteop\ operations, merge,
or collapse steps.  The action for deallocation is a straightforward
push onto the free list for the segment and assigning $\lambda$
to the nodes's parent pointer.  
 
\subsection{Refusal} \label{refusal}

Operations succeed or fail in a legitimate 2-3 tree depending on
the tree content, but conventional implementations do not encounter
situations of single children, paths not terminating at items, and
so forth.  A node with an only child does not cause an operation to 
fail, since a postcondition of truncation is that the node's key 
is equal to some key of its child.  However a path from the root,
guided by key values, which does not end up at segment $S_1$, 
prevents operation completion (for instance, a path may not
lead to a node in $S_1$ because truncation terminates the path).  
Operations fail in these cases.

Thus an \insertop\ operation fails, yielding a ``full'' response, 
if the insertion path prematurely ends --- although sufficient 
free nodes for insertion may exist.  Another instance of \insertop\ 
failure results when detached nodes in all segments do not 
exist, even if the number of items in the tree is far less than
the tree's capacity.  

\subsection{Background Cleaning}

Operations modified to include truncation, merge and collapse
procedures can convert an illegitimate tree into a legitimate 2-3 tree,
provided a sequence of these operations have an appropriately diverse 
set of key parameters.  Of course, we cannot depend on the good fortune
of operation parameters to stabilize the data structures.  Moreover, 
the modifications described above do not address problems of 
illegitimate free lists.  The remaining tasks of stabilization can
be called ``background cleaning'' of the data structure.  We
describe these tasks first as concurrent activities, and later 
show how they can be integrated into the sequential operations 
on the data structure without increasing operation time complexity.

A straightforward approach to correcting an illegitimate tree
would be a systematic visitation of all nodes reachable from the 
root, applying truncation and merger from lowest to highest
segments.  Such a systematic visitation could be a continual background
activity.  A complication with this approach is that operations
may be applied to the data structure concurrently.  To avoid this
complication, we describe a visitation of the nodes that is easily
interleaved with operations.  Let \locate\ be an internal operation
differing from \findop\ only in its failure response:  instead of 
returning ``missing'' if a key $k$ is not contained in the tree,
the response to $\locate(k)$ will be the smallest key $k'$ such
that $k'>k$ if such $k'$ exists in some tree item, 
otherwise $\locate(k)$ returns ``missing''.  
Many implementations of search trees provide
an operation similar to \locate\ so that enumerations of the tree's
items can be easily programmed.  The implementation of \locate\ 
applies truncation, merge and collapse steps as described above.
The background activity consists of the continual repetition 
of the following:  invoke $\locate(t)$, where $t$ is the 
``current'' key value;  if the response is ``missing'', then 
assign to $t$ the least possible value in the domain of key values,
otherwise assign to $t$ the next greater possible value than 
the key value in the response.  The background activity is thus
a round-robin visitation of the items of the tree.  Observe that
the initial current key value $t$ is unimportant to this activity.     

The remaining issue for background activity is the collection
of orphan nodes that should belong to free lists.  There are 
\textit{pmax} such background activities, one for each segment.
For segment $S_i$ this activity consists of a scan of all 
nodes within $S_i$ in round-robin order.  The scan tests each
node $p$ with an $\intree(p)$ function to determine whether
or not $p$ is contained in the tree;  if $p$ is not
in the active tree, then $p$ is pushed onto the free list
for segment $S_i$.   

The $\intree(p)$ test has a recursive definition.  
If $p$ is the root, then $\intree(p)$ is defined to be \textit{true}.  
Cases where $\intree(p)$ is \textit{false} are:   
if $\parent(p)$ does not have a corresponding relevant
key and child pointer to $p$; if $\segment(p)\geq\segment(\root)$ 
for $p\neq\root$; or  
if $\segment(\parent(p))\neq\segment(p)+1$.
Finally, if none of these cases apply,    
then the definition is recursive:  $\intree(p)=\intree(\parent(p))$.
The worst-case running time for $\intree(p)$ is proportional 
to the height of the root.  In an arbitrary initial state of the 
data structure, $\intree(p)=\textit{true}$ does not imply 
that $p$ is part of the active tree, however,
$\intree(p)=\textit{false}$ does imply that $p$ 
not in the active tree.  After 
the data structure stabilizes to a legitimate 2-3 tree by 
sufficiently many truncation and merge steps, any subsequent
$\intree(p)=\textit{true}$ does imply that $p$ is in the active tree. 

The \intree\ test identifies orphan nodes by a negative response,
but the negative response is also returned for nodes in the 
free list.  Thus \intree\ does not precisely identify those nodes
that should be in the free list but are not currently in the free
list.  Our approach is to force any $p$ for which $\intree(p)$ is
\textit{false} to have detached status (i.e., assigning 
$\parent(p)\leftarrow\lambda$) and then moving it to the front
of the free list for its segment.  So that the number of nodes
in a free list are not decreased, the free list has a bidirectional  
pointer implementation:  moving any node in the free list to
the front takes $O(1)$ time and the number of nodes in the free
list is unchanged.

\subsection{Cleaning in Operations}

The 2-3 tree is a passive data structure with no
independent, autonomous processes to perform background
cleaning activities.  Therefore, each operation on the
data structure contributes some processing to the 
background activities.  Another way to state this is
that a sequence of data structure operations simulates
background processes in addition to normal work of
the operations.  Seen as an ongoing simulation, the
various background processes require some 
state information that is saved when the 
simulation is suspended between operations,  
and restored at the start of each operation
to continue the simulation.  The state information
for such suspended background activity results in
the following auxiliary variables:  \curkey\ 
contains the key value used in the \locate\ 
traversal of the tree's nodes;  $\curnode[i]$,
for $0<i\leq\textit{pmax}$, is the current node
location in segment $S_i$ for the round-robin
collection of free nodes;  and \scount\ is 
an integer counter used to control the 
rate of the simulation.

Each data structure operation (\insertop, \deleteop,
\findop) contributes to cleaning by invoking 
\locate\ twice and attempting eleven free node
collections (that is, subjecting eleven nodes to 
the \intree\ test and moving nodes not in the 
tree to the free list).  Each \locate\ uses 
and increments \curkey, and each collection attempt
advances the round-robin \curnode\ location.  
Not all of the eleven collection attempts occur
in the same segment, nor does each operation 
repeat the same selection for the collection attempts:
the value of \scount\ determines, for each attempt,
which segment is chosen.  For each attempt, the 
segment choice is $S_i$ where $i$ is the largest positive
value such that $\scount\bmod 2^{i-1} = 0$, with 
$\scount\leftarrow(\scount + 1)\bmod K$ executed
after each collection attempt.  
\begin{lemma} \label{alloc-rate} \emph{
For any sequence of $k$ data structure operations, 
for $0<i\leq\textit{pmax}$,
at least $\lfloor 11k / 2^i\rfloor$ collection attempts occur in 
segment $S_i$.
}\end{lemma}
\begin{proof}
The $k$ data structure operations generate $11k$ collection attempts,
and for each attempt finding \scount\ odd, segment $S_1$ is chosen.
Thus either $\lfloor 11k/2\rfloor$ or $\lceil 11k/2\rceil$ collection
attempts occur at $S_1$, and the remaining collection attempts
occur in some $S_i$, $1<i\leq\textit{pmax}$.  A similar observation
holds for $S_2$, and recursively, to verify the lemma. 
\end{proof}
\section{Verification} \label{verification}
\begin{theorem}  \emph{
The construction of \S\ref{modops} satisfies availability.
}\end{theorem}
\begin{proof}
The proof is a straightforward verification that
all operations (\insertop, \deleteop, \findop) consult
and modify only the active tree, and background activities
do not remove items from the active tree.  Thus the 
content of the search tree is defined as the set of items
contained in the $S_1$ nodes of the active tree.
\end{proof}  

\paragraph{Notation.}  Let $\bar{n}$ be the number of items
in the initial active tree and let $\bar{m}$ be the number
of nodes in the initial base tree.  Let $\bar{n}_i$ be the number of
active tree nodes in segment $S_i$; we also denote $\bar{n}$
by $\bar{n}_0$.  If the initial active tree is a 2-3 tree, 
it follows that $\bar{n}_i\leq\bar{n}_{i-1}/2$, so 
$\bar{n}_i\leq\bar{n}/2^{i-1}$.  The overbar notation refers
to the initial active tree, and we remove the overbar when
counting nodes at subsequent points in a history of operations.
Thus $n_i$ is the number of active tree nodes in $S_i$ at
a specified state, $\bar{f}_i$ is the initial size of the
free chain in $S_i$, and $f_i$ is the size of the free
chain at a specified state. To refine the node counts, let $n^r_i$ 
be the number of active tree nodes with $r$ children in 
segment $i$, $i>1$, and be the number of tree nodes with
$r$ relevant keys for $i=1$.  Thus $\bar{n}^3_2$ is the 
number of active tree nodes with three children in $S_2$
at the initial state.

Call any state $\sigma$ a \emph{normal state} if the base tree 
$T_\sigma$ equals the active tree of $\sigma$ and this 
active tree is a 2-3.  A state $\sigma$ is \emph{safe}
if it is a normal state and for every segment $S_i$, either
$f_i + n_i = |S_i|$ or $f_i\geq 2n_i$.

\begin{lemma} \label{converge-zero} \emph{
Any data structure operation applied to a normal state
results in a normal state.
}\end{lemma}

\begin{lemma} \label{converge-one} \emph{
Starting from any initial state of the data structure,
the active tree equals the base tree, and the active 
tree is a 2-3 tree with $n=O(\bar{m})$, 
within $O(\bar{m})$ operations of any history.
}\end{lemma}
\begin{proof}
The proof relies on arguments using truncation and background cleaning
to show that $O(\bar{m})$ operations are  
sufficient to visit, check, and correct all nodes of the 
initial base tree (including the possibility that successful \insertop\ 
operations increase the size of the base tree during this sequence 
of $O(\bar{m})$ operations).
\end{proof}

Texts describing 2-3 trees or B-trees observe that the frequency
of node splits and merges decreases geometrically with tree height.
Such observations are simple to verify given an initially empty
tree and then considering worst-case sequences of operations.   
For our purposes, this observation should be enhanced to consider
an initially non-empty tree.  

\begin{lemma} \label{split-gro} \emph{
In a sequence of $t$ operations, $t>\bar{n}_i$, at most
$\bar{n}_i+\lceil(t-\bar{n}_i)/2^i\rceil$ node splits occur 
in segment $S_i$. 
}\end{lemma}
\begin{proof}
In the worst case, each of the $\bar{n}_i$ tree nodes in $S_i$ 
have three children each, so $\bar{n}_i$ of the $t$ operations
can split these initially present nodes.  The usual maximum rate of 
splitting is once per $2^i$ \insertop\ operations, and the lemma
states both of these observations.
\end{proof} 

\begin{lemma} \label{converge-two} \emph{
Any sequence of $\bar{n}$ operations applied to an 
initially safe state results in a safe state, and
no \insertop\ operation fails during this 
sequence of operations unless $n>K$. 
}\end{lemma}
\begin{proof}
To reason about progress over the course of a sequence of 
operations on the data structure, a type of variant function
is useful.  We use for each segment $S_i$ a four tuple 
$\langle n^3_i\comma\ n^2_i\comma\ f_i\comma\ c_i\rangle$, where $n^3_i$, 
$n^2_i$, and $f_i$ are defined above, 
and $c_i$ is the number of collection attempts that
have previously occurred in $S_i$ (from the initial state
to the current state).  The evolution of this tuple for
different types of operations is summarized as follows.
\begin{eqnarray*}
\langle n^3_i\comma n^2_i\comma\ f_i\comma\ c_i \rangle
	&\stackrel{(a)}{\longrightarrow}&
	\langle n^3_i-1\comma\ n^2_i+2\comma\ f_i-1
	\comma\ c_i+\delta_i \rangle\\
\langle n^3_i\comma n^2_i\comma\ f_i\comma\ c_i \rangle
	&\stackrel{(b)}{\longrightarrow}&
	\langle n^3_i+1\comma\ n^2_i-1\comma\ f_i\comma\ c_i+\delta_i \rangle\\
\langle n^3_i\comma n^2_i\comma\ f_i\comma\ c_i \rangle
	&\stackrel{(c)}{\longrightarrow}&
	\langle n^3_i\comma\ n^2_i\comma\ f_i\comma\ c_i+\delta_i \rangle \\
\langle n^3_i\comma n^2_i\comma\ f_i\comma\ c_i \rangle
	&\stackrel{(d)}{\longrightarrow}&
	\langle n^3_i+1\comma\ n^2_i-2\comma\ f_i+1\comma\ 
			c_i+\delta_i \rangle \\
\langle n^3_i\comma n^2_i\comma\ f_i\comma\ c_i \rangle
	&\stackrel{(e)}{\longrightarrow}&
	\langle n^3_i-1\comma\ n^2_i+1\comma\ f_i\comma\ c_i+\delta_i \rangle 
\end{eqnarray*}
In this table, $\delta_i$ represents a nondeterministic
number of collection attempts in $S_i$ (ranging between zero and five)
addressed by Lemma \ref{alloc-rate} for a single operation.
The types of operations in the table are
\itp{a} a node split, \itp{b} a key insertion without
a split, \itp{c} a \findop, unsuccessful \insertop\ or \deleteop,
or a successful \insertop\ or \deleteop\ affecting segments
below $S_i$ but making no change at $S_i$, \itp{d} a node
merge, and \itp{e} removal of a key without a merge.   
Only the transition \itp{a} increases the number of tree
nodes in a segment, and we introduce simpler notation
for this, summing $n^3_i$ and $n^2_i$ to make a triple: 
\begin{eqnarray*}
\langle n_i\comma\ f_i\comma\ c_i \rangle
	&\stackrel{(a)}{\longrightarrow}&
	\langle n_i+1\comma\ f_i-1\comma\ c_i+\delta_i \rangle 
\end{eqnarray*}
Although the additive factor $\delta_i$ in the table above
is indeterminate, Lemma \ref{alloc-rate} does provide a lower
bound for a sequence of operations.  Since our goal is to 
establish sufficient free chain size, we consider the worst
case sequence of operation to deplete a free chain, namely
a sequence of type \itp{a} transitions.  For a sequence of $t$
data structure operations starting from the initial state, 
with $t>\bar{n}$, the result of the sequence satisfies 
\begin{eqnarray*}
\langle \bar{n}_i\comma\ f_i\comma\ 0 \rangle
	&\stackrel{r_i\times(a)}{\longrightarrow}&
	\langle \bar{n}_i+r_i\comma\ f_i-r_i\comma\ s_i \rangle
\end{eqnarray*}
where $r_i$, the number of type \itp{a} transitions in 
segment $S_i$ satisfies $r_i\leq\bar{n}_i+\lceil(t-\bar{n}_i)/2^i\rceil$ 
by Lemma \ref{split-gro}, and $s_i\geq\lfloor 11t / 2^i\rfloor$ 
by Lemma \ref{alloc-rate}.  Because we require bounds only 
we can write $r_i\approx(t+\bar{n}_i)/2^i$ and $s_i\approx 11t/2^i$.
Approximate bounds are expressed by
\begin{eqnarray*}
\langle \bar{n}_i\comma\ f_i\comma\ 0 \rangle
	&\stackrel{r_i\times(a)}{\longrightarrow}&
	\langle \bar{n}_i+(t+\bar{n}_i)/2^i\comma\  
	        f_i - (t+\bar{n}_i)/2^i\comma\ 11t/2^i \rangle 
\end{eqnarray*}
Also, the initial value $\bar{n}_i$ lies in the range
$[\bar{n}/3^i,\bar{n}/2^i]$ by the definition of a 2-3 tree, 
so a conservative bound on the free chain size is given by
\begin{eqnarray*}
\langle \bar{n}/2^i\comma\ f_i\comma\ 0 \rangle
	&\stackrel{r_i\times(a)}{\longrightarrow}&
	\langle \bar{n}/2^i+(t+\bar{n}/2^i)/2^i\comma\  
	        f_i - (t+\bar{n}/2^i)/2^i\comma\ 11t/2^i \rangle 
\end{eqnarray*}
We now distinguish between two cases, 
$\bar{n}=O(K)$ and $\bar{n}=o(K)$.
For the case $\bar{n}=O(K)$ recall that each segment $S_i$ 
has approximately half of the elements of 
$S_{i-1}$, with $S_1$ having about $K/2$ elements 
(so that, if each element is a node with two items,
the capacity $K$ has been attained).
It follows that within $O(\bar{n})$ operations,
every element of every segment undergoes a collection attempt.
Thereafter, each element of $S_i$ is either on the free
chain for $S_i$ or is a node in the active tree.  In such a state,
an \insertop\ operation fails only if the active tree contains
at least $K$ items, which establishes the theorem's conclusion.
\par
For the case $\bar{n}=o(K)$, we examine a history of $\bar{n}$ 
operations ($t=\bar{n}$).  For bounding the free chain
size, we then have
\begin{eqnarray*}
\langle \bar{n}/2^i\comma\ f_i\comma\ 0 \rangle
	&\stackrel{r_i\times(a)}{\longrightarrow}&
	\langle \bar{n}/2^i+(\bar{n}+\bar{n}/2^i)/2^i\comma\  
	        f_i - (\bar{n}+\bar{n}/2^i)/2^i\comma\ 11\bar{n}/2^i \rangle 
\end{eqnarray*}
An overestimate of the count of nodes and size of free chain is
obtained by the substitution of $\bar{n}$ for $\bar{n}/2^i$, given by
\begin{eqnarray*}
\langle \bar{n}/2^i\comma\ f_i\comma\ 0 \rangle
	&\stackrel{r_i\times(a)}{\longrightarrow}&
	\langle 3\bar{n}/2^i\comma\  
	        f_i - 2\bar{n}/2^i\comma\ 11\bar{n}/2^i \rangle 
\end{eqnarray*}
Thus we see $11\bar{n}/2^i$
collection attempts in $S_i$ exceeds the number of active 
tree nodes by at least $8\bar{n}/2^i$;  this implies that 
after the $\bar{n}$ operations, at least $8\bar{n}/2^i$ 
collection attempts occur outside of active tree nodes.  
Of course, some or all of these collection attempts could
apply to elements already in the free chain.  So, while not
every collection attempt outside the active tree results
in an increase in the free chain size, the $11\bar{n}/2^i$
collection attempts do ensure a free chain size of at 
least $8\bar{n}/2^i$, less any elements consumed by splits
during the period of these collection attempts.  Since 
the number of elements consumed is $2\bar{n}/2^i$ during
this period, it follows that the free chain size is 
at least $6\bar{n}/2^i$.  Thus if $f_i=2\bar{n}/2^i$ 
(the minimum needed to permit all the splits), then 
the size of the free chain after $\bar{n}$ operations
is at least $6\bar{n}/2^i$ in segment $S_i$.  

A conclusion of this analysis is that $\bar{n}$ operations
at most triple the number of tree nodes in $S_i$, 
while multiplying the free list size by a factor of six.
The analysis also shows that $f_i\geq2\bar{n}_i$ is sufficient
to supply all node allocation.  Hence, the result of 
applying $\bar{n}$ operations supplies sufficiently many 
free list elements for a subsequent sequence of $\bar{n}$ 
operations (because $6\bar{n}/2^i$ is twice $3\bar{n}/2^i$).   
\end{proof}

\begin{lemma} \label{converge-three} \emph{
Any sequence of $\bar{n}$ operations applied to an 
initially normal state results in a safe state. 
}\end{lemma}
\begin{proof}
The analysis presented in the proof of Lemma \ref{converge-two}
holds for purposes of bounding the free chain size even when
operations are not of type \itp{a}, which shows that after
$\bar{n}$ operations every free chain size either includes
all non-tree nodes or is double the number of free chain nodes.
\end{proof}

\begin{theorem} \emph{
The construction of \S\ref{modops} is stabilizing with $O(\bar{m})$
stabilization time.
}\end{theorem}
\begin{proof}
Lemmas \ref{converge-zero} and \ref{converge-two} show stability
for a tree in safe state.  Lemma \ref{converge-one} states that
$O(\bar{m})$ operations suffice to reach a normal state, and 
Lemma \ref{converge-three} implies that within $O(\bar{m})$ 
subsequent operations, the state is safe. 
\end{proof}
\section{Discussion} \label{conclusion}

The construction presented here shows that goals of availability 
and stabilization are achievable for a search tree.  The solution
is adaptive, however the adaptive stabilization period is related
to the number of nodes of the initial active tree rather than
the number of items (in \cite{HM00} the adaptivity is 
linear in the size of the initial active heap).  This seems 
unavoidable, since the active tree could initially have only
one item, but $\textit{pmax}=O(\lg K)$ nodes, and it is not
possible to recognize the active tree and truncate it by 
$O(1)$ operations that are each limited to $O(\lg K)$ running time.   

An important issue not addressed in this paper is limiting the 
amount of data lost due to a transient fault to be proportional
to the scope of that fault.  If the root of the tree is lost, then
all of the items of the data structure are lost by our construction.
So, in the worst case, damage to a single node can lead to loss of
all data.  At the other extreme, damage to a leaf node only results
in loss of the data at the leaf.   If minimizing data loss is an 
important goal, then data could be secured in higher level segments 
of the tree using replication techniques \cite{R89,KP95,AB96} to 
reduce the probability of loss by a transient fault.  The degree of
replication could be made proportional to the height of the node.
If fault probability distributions have location independence, then
it could be that the probability of losing data is roughly uniform for any 
node (least likely at higher levels due to replication, but with larger
impact when it does occur).  Using such a replication would have added
storage cost and also a cost in operation times, since each operation 
would verify sufficient consistency among replicas.


\begin{thebibliography}{99}
\bibitem{AD97a} Y Afek and S Dolev.  Local stabilizer.
In \emph{Proceedings of the 5th Israeli Symposium on
Theory of Computing and Systems}, pp. 74-84, 1997.
\bibitem{AHU74} AV Aho, JE Hopcroft, JD Ullman. 
\emph{The Design and Analysis of Computer Algorithms}, 1974,
Addison-Wesley.
\bibitem{DH97} S Dolev and T Herman.
Superstabilizing protocols for dynamic distributed systems. 
\emph{Chicago Journal of Theoretical Computer Science}, 3(4), 1997. 
\bibitem{FMRT96} MJ Fischer, S Moran, S Rudich, and G Taubenfeld.
The wakeup problem. \emph{SIAM Journal on Computing}, 25:1332-1357, 1996.  
\bibitem{HW94} M Herlihy, JM Wing.
Specifying graceful degradation in distributed systems. 
\emph{ACM TODS}, 19(4):586-625(1994).
\bibitem{H99} T Herman.  Superstabilizing mutual exclusion. 
\emph{Distributed Computing}, 13(1):1-17, 2000.
\bibitem{HM00} T Herman, T Masuzawa.
Available stabilizing heaps.  To appear in 
\emph{Information Processing Letters}, 2000. 
\bibitem{KS97} S Kutten, B Patt-Shamir.  Time-adaptive 
self stabilization.  In \emph{Proceedings of the 16th Annual ACM Symposium
on the Principles of Distributed Computing}, pp. 149-158, 1997.
\bibitem{S93c} M Schneider.  Self-stabilization. 
\emph{ACM Computing Surveys}, 25:45-67, 1993.
\bibitem{UKMF97} E Ueda, Y Katayama, T Masuzawa, and H Fujiwara.
A latency-optimal superstabilizing mutual exclusion protocol.
In \emph{WSS'97 Proceedings of the Third Workshop on Self-Stabilizing
Systems}, pp. 110-124, 1997.
\bibitem{R89} MO Rabin.
Efficient dispersal of information for security, load balancing, and fault
tolerance. \emph{Journal of the ACM} 38:335-348, 1989.
\bibitem{AB96} Y Aumann, MA Bender.
Fault tolerant data structures.  In
\emph{Proceedings of the 37th Annual Symposium on Foundations of
Computer Science (FOCS '96)}, pp. 580-589, 1996.
\bibitem{KP95} S Kutten, D Peleg. Fault-local mending.
In \emph{Proceedings of the 14th Annual ACM Symposium
on the Principles of Distributed Computing}, pp. 20-27, 1995.
\end{thebibliography}
\end{document}